\documentclass[12pt,preprint]{aastex}
\newcommand{\mincir}{\raise
-2.truept\hbox{\rlap{\hbox{$\sim$}}\raise5.truept\hbox{$<$}\ }}
\newcommand{\magcir}{\raise
-2.truept\hbox{\rlap{\hbox{$\sim$}}\raise5.truept\hbox{$>$}\ }}
\newcommand{\minmag}{\raise
-2.truept\hbox{\rlap{\hbox{$<$}}\raise6.truept\hbox{$<$}\ }}

\begin{document}

\title{
Environmental Effects of Dark Matter Haloes: 
The Clustering-Substructure relation of Group-size Haloes}
\author{N\'estor Espino-Briones\altaffilmark{1}, 
Manolis Plionis\altaffilmark{2,1} and
Cinthia Ragone-Figueroa\altaffilmark{3,4}}
\altaffiltext{1}{Instituto Nacional de Astrof\'isca \'Optica y Electr\'onica,
 AP 51 y 216, 72000, Puebla, M\'exico}
\altaffiltext{2}{Institute of Astronomy \& Astrophysics, National Observatory of
 Athens, Palaia Penteli 152 36, Athens Greece}
\altaffiltext{3}{Grupo IATE-Observatorio Astron\'omico, Laprida 854, C\'ordoba Argentina}
\altaffiltext{4}{Consejo de Investigaciones Cient\'ificas y T\'ecnicas de la 
R\'epublica Argentina, C\'ordoba, Argentina}

\begin{abstract}
We estimate the two-point correlation function of dark matter haloes, 
with masses $\ge 10^{13}\;h^{-1} \;M_{\odot}$, that have or not significant 
substructure. 
The haloes are identified with a friends of friends algorithm in
a large $\Lambda$CDM simulation at two redshift snapshots ($z=0.0$ and 
$1.0$), while halo substructure is determined using an observationally driven
method. We find in both epochs a clear and significant
signal by which haloes with substructure are more clustered than those with
no-substructure. This is true for all the considered halo mass ranges, 
although for the highest halo masses the signal is noisy and present
only out to $\sim 20 h^{-1}$ Mpc. There is
also a smooth increase of the halo correlation length with increasing
amplitude of the halo substructure.
We also find that substructured haloes are typically located in high-density
large-scale environments, while the opposite is true for
non-substructured haloes.
If the haloes found in high-density regions have a relatively earlier formation
time, as suggested by recent works, then they do indeed have more time
to cluster than haloes, of a similar mass, which form later in 
the low-density regions. 
In such a case one would have naively expected that the former (earlier formed) haloes 
would typically be dynamically more relaxed than the latter (later formed). 
However, the higher merging and interaction rate,
expected in high-density regions, could disrupt their relatively relaxed 
dynamical state and thus be the cause for the higher fraction of
haloes with substructure found in such regions.
\end{abstract}


\section{Introduction}
In the current cold dark matter picture, structures form
via gravitational amplifications of initial density fluctuations.  
The dark matter (DM) haloes 
form hierarchically 
by the aggregation, via gravitational interactions, of 
small collapsed structures that merge to form larger ones, eventually
forming clusters of galaxies.

Observationally, galaxy properties show a significant variation 
depending on their local and large-scale environment
(eg. Dressler 1980; G\'omez et al. 2003; Boselli \& Gavazzi 2006). 
Since galaxies form within DM haloes, the ``parent'' halo properties 
could influence or even determine the galaxy properties.
Therefore, many theoretical studies have investigated in detail the
inter-relation of the DM halo properties,
among which the halo formation time, their assembly histories,
their central concentration, their clustering and their location
in the cosmic web. 
The first studies (eg. Lemson \& Kauffmann 1999) found no
significant environmental dependence of the various DM halo
properties. However, a re-interpretation of their results and a wealth
of new studies point in the opposite direction.

Halo formation time has been found to correlate with halo
concentration and with the number of sub-haloes
(eg. Navarro, Frenk \& White 1996; Jing 2000; Bullock
et al. 2001; Wechsler et al. 2002; Zhao et al. 2003; Zentner et al. 2005).
Sheth \& Tormen (2004) showed that haloes in
dense regions form at slightly earlier times than similar mass haloes
in lower density regions, a result confirmed by Zhu et
al. (2006) and Harker et al. (2006) for massive haloes as well. 
Gao et al. (2004) and Gao \& White (2007) extended the halo ``assembly bias" 
investigation by considering different parent halo properties, among which
concentration, spin and subhaloes mass fraction.

Furthermore, the amplitude of the DM halo two-point correlation function 
has been found to depend strongly on the halo formation 
time. Relatively low-mass haloes ($\mincir 10^{13} h^{-1} M_{\odot}$),
that assembled at high redshifts, are more clustered than 
those that assembled more recently (Gao, Springel \& White 2005; Wechsler et
al. 2006; Wang et al. 2007; see also Berlind et al. 2006), an effect which 
strengthens with decreasing halo mass and with decreasing halo separations.
However, Wetzel et al. (2006) and Jing, Suto \& Mo (2007) found that
very massive haloes, which have formed recently, appear to cluster more
strongly than older haloes of the same mass, which is the opposite
than what has been found for lower mass haloes (see also Wechsler et al. 2006).

Regarding the dependence of halo clustering on subhalo occupation number,
Wechsler et al. (2006), studying only large host haloes, found 
close pairs of haloes having above-average number of satellites.
Gao \& White (2007), using the Millennium simulation (Springel et al. 2005), 
demonstrated that haloes with a relatively large mass fraction in sub-haloes
are more strongly clustered, confirming Wechsler et al. (2006). 
Thus halo occupation by galaxies could be a function of environment.
This appears to be in agreement with some observational cluster studies,
to the extent to which halo-occupation is related to the halo dynamical state.
For example, dynamically active clusters (having significant substructure) 
have been 
found to be preferentially located in dense large-scale environments 
(Schuecker et al. 2001; Plionis \& Basilakos 2002), while they are also 
more clustered with respect to clusters with weak or no substructure 
(Plionis \& Basilakos 2002).
  

Further on the observational side, Blanton et al. (2006) 
found a slight dependence between galaxy colours and measures of the 
surrounding environment on various scales.
Berlind et al. (2006), searching for a
dependence of the SDSS galaxy groups clustering 
on a second parameter, independent of mass, 
found a correlation between the large scale bias of massive groups 
and their central galaxy $(g-r)$ colour as well as hints of a possible connection between 
the later and DM halo age.
Also Croton, Gao \& White (2007) studying the relation between the clustering of 
DM haloes and their assembly history, using a galaxy formation model applied
in the Millennium simulation,
found that the colour of the central galaxy in a DM halo, of a given 
mass, depends on the halo's environment.

All these results put into question the simplest form of
the excursion-set formalism for
galaxy clustering (Bond et al. 1991; Lacey \& Cole 1993), 
which predicts that the properties of galaxies,
forming within the DM haloes, are functions only of the 
DM halo mass, and not of other properties like its environment.
Nevertheless, more elaborate schemes of the excursion set formalism
have been recently proposed that allow more
faithful descriptions (Sandvik et all. 2007, Zentner 2007). 


In this letter we address a related issue; ie., the relation between DM
halo clustering, the halo dynamical state and its large-scale environment.
Note that we do not use the common
approach of counting bound sub-haloes, within parent haloes, as an
indicator of substructure, but rather
an observationally driven estimator of the halo dynamical state
 which is based on the dynamical measure of halo substructure
according to Dressler \& Shectman (1988).
Once the halo dynamical state is determined we
compute the two-point correlation function for haloes 
with and without substructure and we also characterize their local environment.

\section{Numerical Simulation \& Methodology}
The numerical simulation used in this work was performed using the 
GADGET2 code (Springel 2005) with dark matter only. 
We use a ${\rm \Lambda CDM}$ cosmology
with $\Omega_m=0.3$, $\Omega_\Lambda=0.7$, 
$\sigma_8=0.9$, $h=0.72$, where $\Omega_m$ and  $\Omega_\Lambda$ are the 
present day matter and vacuum energy densities in units 
of the critical density,
$\sigma_8$ is the present linear rms amplitude of mass fluctuation in 
spheres of 8 $h^{-1}$ Mpc and $h$ is the Hubble parameter in units of 
$100 \rm{km~s^{-1} Mpc^{-1}}$. 
The simulation was run in a cube of size $L=500 \; h^{-1}$ Mpc, using $512^3$ 
particles. The particle mass is $\sim 7.7\times10^{10} \;h^{-1}\; 
M_{\odot}$ and the force softening length is $\epsilon = 100 \;h^{-1}$ kpc.

The DM haloes were identified using a friends-of-friends (FOF) algorithm with a linking length
$l=0.17$ times the mean inter-particle separation. 
Given the purpose of this work, we only use haloes with at least 130
particles, ie., with masses greater than $10^{13} M_{\odot}$, which
results in a sample of $\sim 58000$ haloes at $z=0$ and $\sim 32400$ at $z=1$.
Note that the halo finding algorithm used provides unique haloes that are
not sub-haloes of any other ``parent'' halo. 

Since we wish to investigate the possible correlation between
clustering and the dynamical state of DM haloes, we 
use the Dressler \& Shectman (1988) algorithm to
estimate the amount of halo substructure. Details for a
recent application of this method to simulation data can be found in
 Ragone-Figueroa \& Plionis (2007).
Briefly, this method determines the 
mean local velocity, $\langle \bf v_{\rm loc} \rangle$, 
and the local velocity dispersion, $\sigma_{\rm loc}$, of the nearest $n$ 
neighbours from each halo particle $i$ 
and compares them with the mean velocity, $\langle {\bf V} \rangle$, 
and the velocity dispersion, $\sigma$, of the whole halo of $N$ particles, 
defining the following measure:
\begin{equation}
\delta_i^2 = \frac{n}{\sigma}[(\langle \bf v_{\rm loc}\rangle- \langle
  \bf V\rangle)^2+(\sigma_{\rm loc}-\sigma)^2] \;.
\end{equation}
Then the  quantification of the halo substructure is given by the
so-called  $\Delta$-{\em deviation}, which is the sum of the
individual $\delta_i$'s over all halo particles $N$:
$\Delta = \sum_{N} \delta_i /N$.
The larger the $\Delta$-deviation the stronger is the halo
substructure.
This statistic depends on the number of nearest neighbours $n$ which is 
used in the analysis (eg., Knebe \& M\"uller 1999)
and on the number of particles used to
resolve a halo as well. Such resolution effects were studied in detail
in Ragone-Figueroa \& Plionis (2007), who  
found a monotonic increase of $\langle \Delta \rangle$ 
with the halo mass which is clearly due to resolution effects. 
However, they also found that within relatively small
halo mass intervals, the corresponding sorted $\Delta$-deviation 
distribution can be used effectively to separate substructured from 
non-substructured haloes, by dividing them in those having 
$\Delta$-deviation above and below the corresponding median or some 
quantile of the $\Delta$-deviation distribution within each halo mass interval.

Therefore, the total halo sample was divided 
in 3 mass ranges and the analysis was performed in each individual
subsample and for each of the two redshift snapshots ($z=0$ and $1.0$).
These halo mass ranges are:

\noindent
(a) $10^{13} \; h^{-1} M_{\odot} \le M< 3\times10^{13}h^{-1} M_\odot$, 

\noindent
(b) $3\times10^{13}\; h^{-1} M _{\odot}\le M <10^{14} \; h^{-1} M_\odot$, 

\noindent
(c) $M\ge 10^{14} \;h^{-1} M_\odot$.

\noindent
We compute the DM halo two-point correlation function, $\xi (r)$,
separately for the substructured haloes (dynamically active) and 
non-substructured haloes (virialized), within each mass range, and to
emphasize their possible clustering differences
we separate haloes having $\Delta$-deviation values larger than the 67\% and 
lower than the 33\% quantile of the $\Delta$-deviation frequency distribution function.

The measured halo two-point correlation function is also fitted to a power-law:
$\xi (r)=\left(r/r_0\right)^\gamma$,  in the range $4\mincir r \mincir
30 \; h^{-1}$ Mpc, using a $\chi^2$ minimization procedure.
Furthermore, we estimate the halo mass one-point over-density 
distribution function, dividing the simulation box in grid-cells of 
5 $h^{-1}$ Mpc size and compare the total DM mass (using haloes with
$M>10^{13} h^{-1} M_{\odot}$) within each cell, $M_i$, to its mean
value, $\bar{M}$, ie., estimating: $\delta M/M \equiv (M_i-\bar{M})/\bar{M}$. 
We then assign to each halo the $\delta M/M$
value that corresponds to the grid-cell in which it is located and derive
the corresponding frequency distribution, separately for haloes
that have or not significant substructure. We have verified that 
this local density estimator is equivalent with the traditionally used, estimated in spheres 
centered on each individual halo.
 
\section{Results and Discussion}

In Figure 1 we show the ratio, $R(r)$, of the two-point correlation functions
of haloes with and without substructure, for the three mass intervals and 
for both redshifts epochs. The errors shown are the
propagated quasi-Poissonian uncertainties, estimated by:
$\delta\xi(r) \simeq \sqrt{[1+\xi(r)]/{\rm DD}(r)}$, where ${\rm DD}(r)$ 
are the halo-halo pairs within separation $r\pm \delta r$.
It is evident that $R(r)$ is significantly less than 1, indicating
that the former haloes indeed have a significantly higher correlation function 
than the latter ones. For $z=0$ we have $R(r)\mincir 0.8$ 
for all $r$'s, while for $z=1$ and for the
most massive haloes this is true for separations $\mincir 20$ $h^{-1}$ Mpc.
Note, however, that the size of our simulation does
not allow us to probe effectively this halo mass range.

In order to assign a probability to the events shown in Figure 1, we use
a Monte-Carlo procedure by which we derive $R(r)$ for 
each of 10$^{6}$ random halo sub-sample pairs 
of the same size as those of our main analysis
(and for each mass range and redshift bin). We then ask
how many times is $R(r)$ consistently less than 1, for all 
$r$'s (or for $r\mincir 20 \; h^{-1}$ Mpc in the case of 
$z=1$ and the higher mass bins), to find that
in all cases the probability is $< 0.0009$.
If we now ask a more restrictive but more accurate
question, ie., how many times would $R(r)$
be systematically lower at the level observed between the
substructured and non-substructured halo sub-samples, then the answer
$<10^{-6}$. 
We therefore conclude that haloes with substructure are significantly 
more clustered than haloes without substructure, locally and in high 
redshifts, in agreement with:

\noindent
(a) the observational results based on clusters of galaxies 
(Plionis \& Basilakos 2002),

\noindent
(b) the DM halo occupation number-clustering correlation (Wechsler et
al. 2006; Gao \& White 2007), and

\noindent
(c) the weak signal found by Wetzel et al. (2006) that haloes which
have undergone a recent major merger (ie., have significant substructure
according to our nomenclature) show a slightly enhanced clustering.

To test whether the dependence of the clustering length, $r_0$, on halo 
substructure is due only to those haloes which experience
``major'' mergers (as suggested by Wetzel et al. 2006)
we have divided our haloes, of each mass range, in 4 equal-number 
sub-samples based on their sorted $\Delta$-deviation values
(corresponding to different substructure amplitudes)
and computed their individual two-point correlation function.
In Figure 2 we present the 
corresponding correlation lengths as a function of $\Delta$ for all 
halo sub-samples (the indicated $\Delta$-deviation value corresponds
to its median value in each of the 4 ranges).
It is evident that the correlation length is growing smoothly and monotonically
with increasing $\Delta$-deviation value, a fact which argues against
the $r_0 - \Delta$ relation being only due to ``major''
mergers. A similar result has been found also for real clusters of galaxies
(Plionis \& Basilakos 2002 - their figure 3). 

Now we ask what could the reason be for such an effect? Could it be that 
haloes having substructure (ie., being dynamically active) 
are located in high density environments (see also 
Ragone-Figueroa \& Plionis 2007) and thus at regions 
of a relatively early halo formation time, while haloes with no 
substructure are
located in low-density regions, ie., at regions of a later halo
formation time (eg. Sheth \& Tormen 2004; Zhu et al. 2006; Harker et al. 2006).
The higher clustering of substructured haloes indeed points in 
such a direction, although one would have expected that earlier formed haloes 
would typically be dynamically
more relaxed than equal-mass later formed ones and would have had
less evident substructure
features. However, the higher merging and interaction rate,
expected in high-density regions, could disrupt the relatively relaxed halo
dynamical state.

To attempt to answer the question posed, we present in Figure 3 
the ratio of the normalized $\delta M/M$ frequency
distributions of the substructured and 
non-substructured haloes (for the two extreme mass ranges,
just for economy of space).
If both type of haloes traced similar overdensities, this ratio 
should have been statistically equivalent to one. However, 
it is evident that it is significantly higher (or lower) 
than 1 at larger (or smaller) overdensities, for all halo mass 
ranges and redshifts.
As we had anticipated, substructured haloes are typically found in 
higher-density regions.

\section{Conclusions}
In this letter we have investigated the relation between DM halo clustering,
halo dynamical state and halo large-scale environment. We identified haloes 
with a FOF algorithm in a dark matter only $\Lambda$CDM simulation and used
haloes with $M\gtrsim10^{13} \;h^{-1} \;M_\odot$, identified at $z=0$ 
and $z=1$. The halo dynamical state was determined by 
measuring the amount of halo substructure using an observationally
driven approach.
We then calculated the two-point correlation function for haloes with high and 
low levels of substructure, finding that the former haloes 
are significantly more clustered than the latter,
while there is also a smooth increase of the halo correlation length
with increasing amplitude of the halo substructure index.

Finally, we find a highly significant signal by which 
haloes with high levels of substructure 
are located typically in higher density regions with respect to haloes with 
low levels of substructure and this could be an explanation 
of our previous results.
The higher clustering of haloes found in
high-density large-scale environments should be expected if
haloes collapse earlier in such regions 
and thus have more
time to evolve and cluster, while their higher-levels of substructure 
should be probably attributed to the higher rate of halo interactions 
and merging which is present in such high-density regions.

\acknowledgments
NE-B is supported by a CONACyT studentship at the INAOE.
MP acknowledges funding by the Mexican Government grant No. CONACyT-2006-49878.
CR-F acknowledges support by the European Commission's ALFA-II 
  programme, through its funding of the Latin-American European Network for
  Astrophysics and Cosmology (LENAC), and funding by the
Consejo de Investigaciones Cient\'{\i}ficas y T\'ecnicas de la Rep\'ublica 
  Argentina (CONICET). 
We thank the referee, Andrew Zentner, for useful suggestions.

\clearpage

\begin{figure}
\plotone{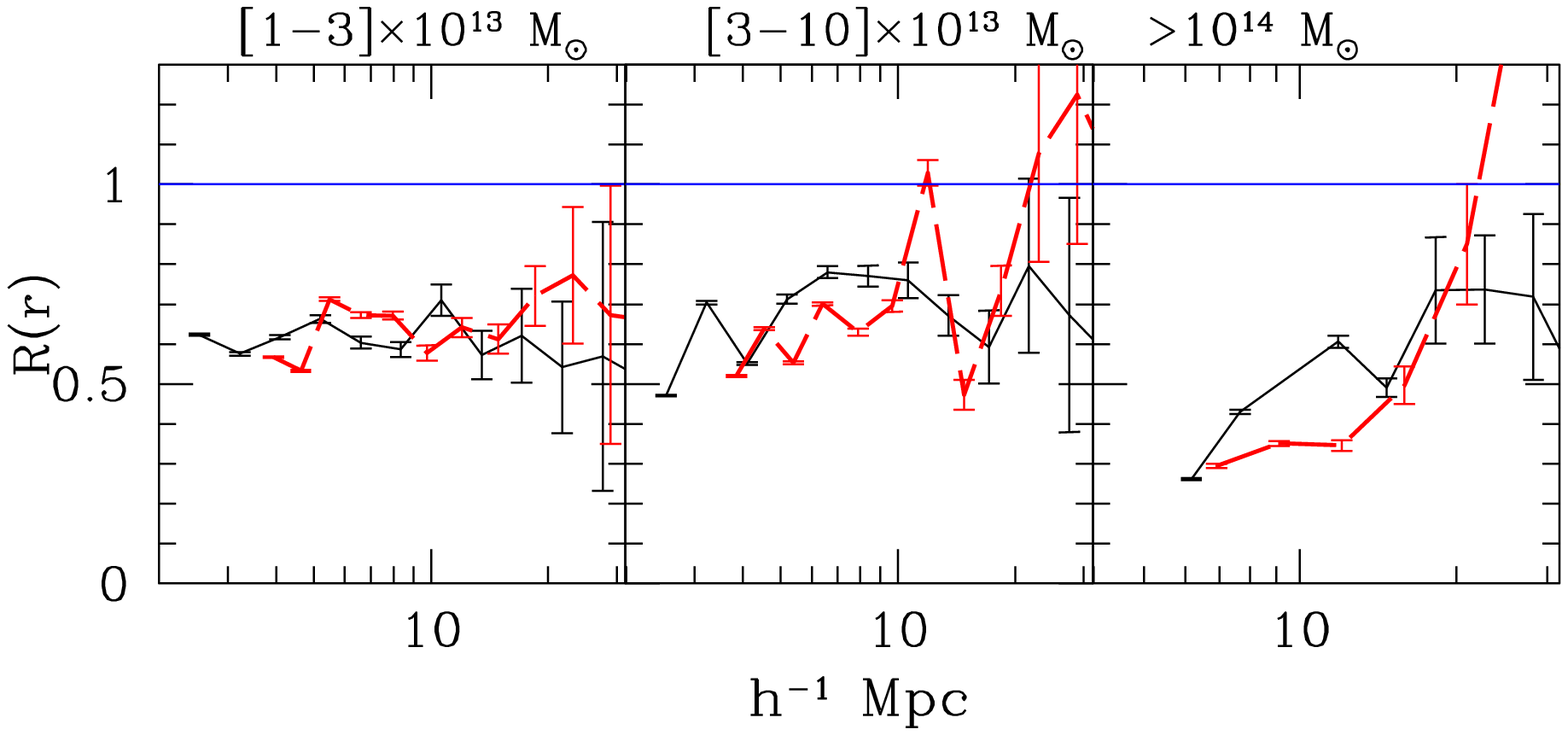}
\caption{The ratio of the 2-point correlation functions
of haloes with and without substructure
for both redshifts epochs. The continuous lines corresponds to $z=0$, 
while the dashed lines to $z=1$. The errors shown are the
propagated quasi-Poissonian $\xi(r)$ uncertainties.}
\label{fig1}
\end{figure}

\clearpage

\begin{figure}
\plotone{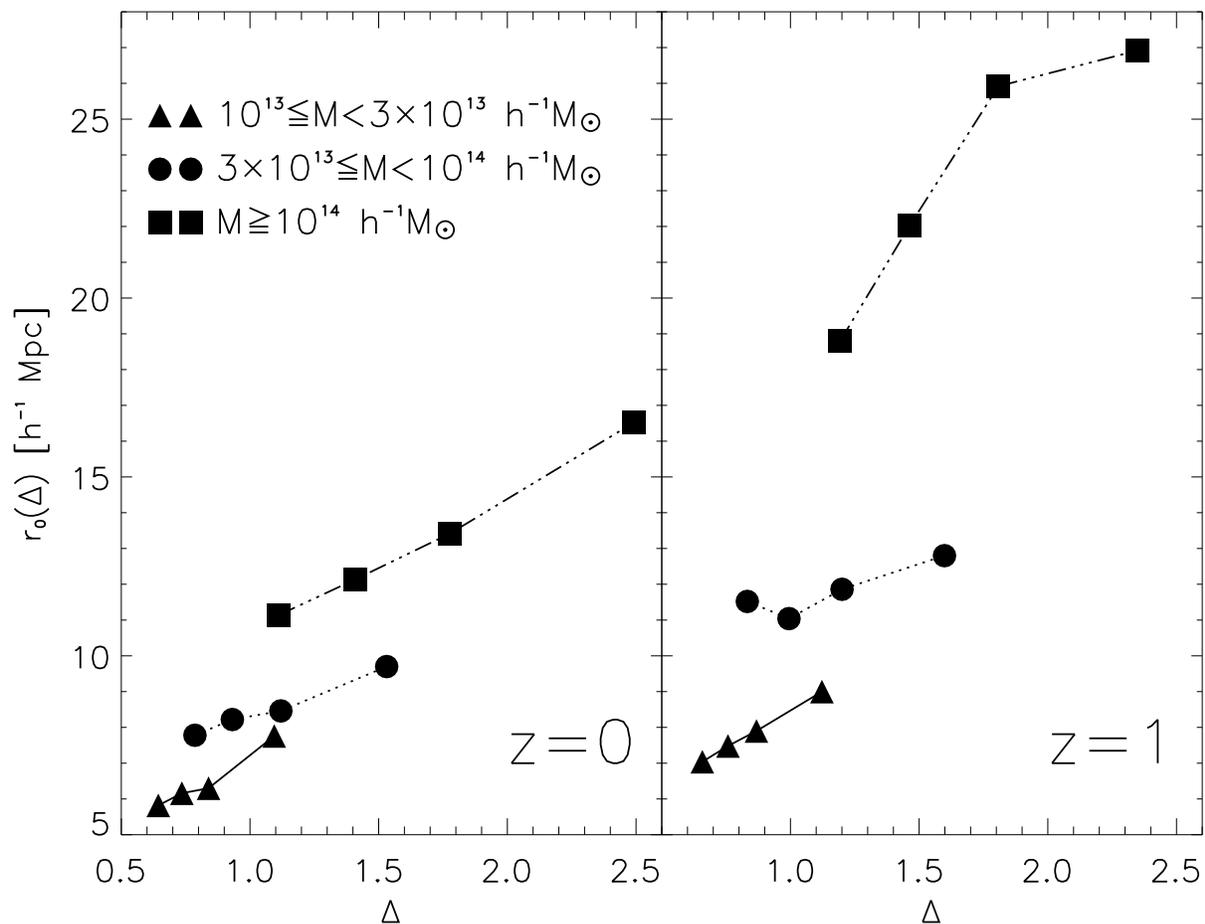}
\caption{Correlation length, $r_0$, as a function of halo $\Delta$-deviation
for the three halo mass ranges and for two redshift snapshots ({\em
left panel:} $z=0$ and {\em right panel:} $z=1$). 
The different halo mass ranges are indicated by the different symbols.}
\end{figure}

\begin{figure}
\plotone{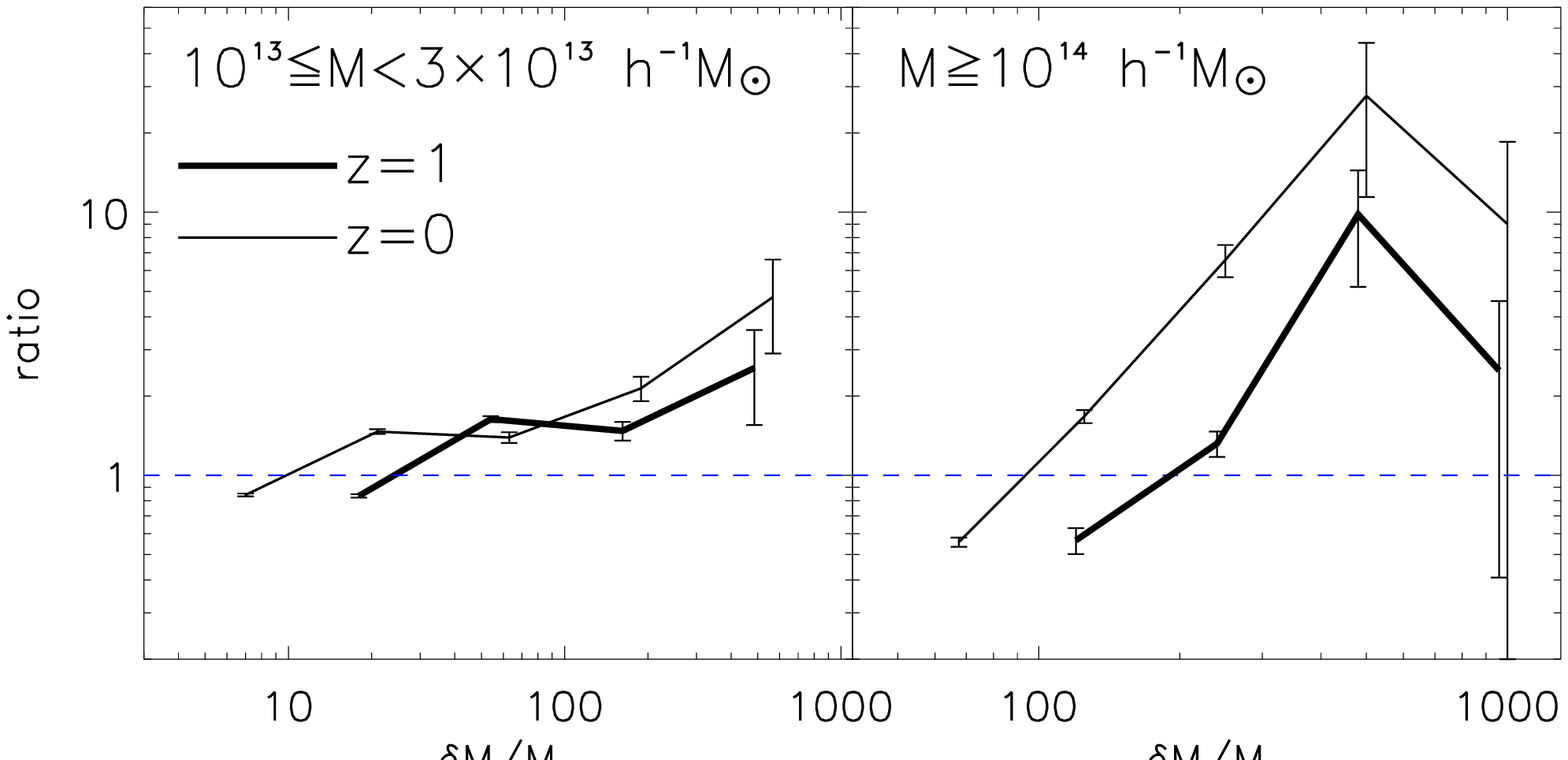}
\caption{The ratio of the normalized $\delta M/M$ frequency 
distributions of substructured and non-substructured haloes. The $\delta M/M$ 
fluctuations have been evaluated on a grid with a 5 $h^{-1}$ Mpc cell-size.
Results are shown for two mass ranges and two redshifts. 
Error-bars are propagated individual Poissonian uncertainties.}
\end{figure}


\begin{thebibliography}{}
\bibitem[]{Be} Berlind, A.A., Kazin, E., Blanton, M.R., Pueblas, S.,
Scoccimarro, R., Hogg, D.W., {\tt astro-ph/0610524} 
\bibitem[]{Bl} Blanton, M., Berlind, A., Hogg, D.W., {\tt astro-ph/0608353}
\bibitem[]{Bon} Bond, J. R., Cole, S., Efstathiou, G., Kaiser, N. 1991, \apj, 379, 440
\bibitem[]{Bo} Boselli, A., Cavazzi, G., 2006, PASP, 118, 517
\bibitem[]{Bu} Bullock, J.S., Kolatt, T.S., 
Sigad, Y.,  Somerville, R.S.,  Kravtsov, A.V.,  Klypin, A.A., Primack,
J.R., \& Dekel, A. 2001, \mnras, 321, 559
\bibitem[]{Cr} Croton, D. J., Gao, L., White S.D.M. 2007, \mnras, 374, 1309
\bibitem[]{Dr1} Dressler A., 1980, ApJ, 236, 351.
\bibitem[]{Dr2} Dressler, A., Shectman, S. A. 1988, \aj, 95, 985
\bibitem[]{Gao2004} Gao, L., White, S.D.M. Jenkins, A., Stoehr, F., Springel, V.
 2004, \mnras, 355, 819
\bibitem[]{Gao} Gao, L., Springel, V., White, S.D.M. 2005, \mnras, 363, L66
\bibitem[]{GaoW} Gao, L. \& White, S.D.M. 2007, \mnras, 377, 5
\bibitem[]{G} G\'omez, P.L., et al., 2003, ApJ, 584, 210
\bibitem[]{Ha} Harker, G., Cole, S., Helly, J., Frenk, C., Jenkins, A.,
2006, MNRAS, 367, 1039
\bibitem[]{Ji} Jing, Y.P., 2000, ApJ, 535, 30
\bibitem[]{Ji1} Jing, Y.P., Suto, Y., Mo, H.J., 2007, ApJ, 657, 664
\bibitem[]{Kn} Knebe A., M\"uller V., 2000, A\&A, 354,761K.
\bibitem[]{La} Lacey, C., Cole, S. 1993, \mnras, 262, 627
\bibitem[]{Le} Lemson G. \& Kauffmann G., 1999, MNRAS, 302, 111
\bibitem[]{} Navarro, J.F., Frenk, C., \& White, S.D.M., 1996, \apj, 462, 563
\bibitem[]{Pl} Plionis, M. \& Basilakos, S., 2002, MNRAS, 329, L47
\bibitem[]{Ra} Ragone-Figueroa, C., Plionis, M. 2007, \mnras, 377, 1785
\bibitem[]{Sandvik} Sandvik, H.B., M\"uller,O., Lee, J., White, S.D.M. 
\mnras, 377, 234S  
\bibitem[]{Sh} Sheth R.K. \& Torment, G. 2004, \mnras, 350, 1385
\bibitem[]{Sch} Schuecker,  P., Boehringer, H., Reiprich, T.H., Feretti, L., 
2001, A\&A, 378, 408
\bibitem[]{Sp} Springel, V., 2005, \mnras, 364, 1105
\bibitem[]{Sp1} Springel, V., et al., 2005, \nat, 435, 639
\bibitem[]{Wa} Wang, H.Y., Mo, H.J., Jing, Y.P., 2007, \mnras, 375, 633
\bibitem[]{We} Wechsler, R.H., Bullock, J.S., Primack, J.R., Kratsov, A.V.
\& Dekel, A. 2002, \apj, 568, 52
\bibitem[]{We1} Wechsler, R. H., Zentner, A. R., 
Bullock, J. S., Kravtsov, A. V., Allgood, B. 2006, \apj, 652, 71
\bibitem[]{Wet} Wetzel, A.R., Cohn, J.D., White, M., Holz, D.E., Warren,
M.S., 2007, ApJ, 656, 139
\bibitem[]{Zentner1} Zentner, A.R., Berlind, A.A., Bullock, J.S., Kravtsov, 
A.V., Wechsler, R.H., 2005, \apj, 624, 505
\bibitem[]{Zentner2} Zentner A.R., 2007, IJMPD in press, {\tt astro-ph/0611454}
\bibitem[]{Zh} Zhao, D.H., Jing, Y.P., Mo, H.J., B\"orner, G., 2003,
ApJ, 597, L9
\bibitem[]{Zhu} Zhu,G., Zheng Z.,  Lin, W.P.,  Jing, Y.P., Kang, X.,
Gao, L., 2006, \apj, 639, L5
\end{thebibliography}
\end{document}